\documentclass[titlepage,aps,titlepage,prd,superscriptaddress,twocolumn,longbibliography]{revtex4-1}
\usepackage[]{graphicx, subcaption, placeins,mathtools,footmisc}
\makeatletter
\renewcommand*{\@fnsymbol}[1]{\ensuremath{\ifcase#1\or *\or \dagger\or \ddagger\or
    \mathsection\or \mathparagraph\or \|\or **\or \dagger\dagger
    \or \ddagger\ddagger \else\@ctrerr\fi}}
\makeatother
\usepackage{newunicodechar}
\usepackage{libertine}
\DeclareRobustCommand{\okina}{%
  \raisebox{\dimexpr\fontcharht\font`A-\height}{%
    \scalebox{0.8}{`}%
  }%
}
\newunicodechar{ʻ}{\okina}

\begin{document}
\bibliographystyle{apsrev4-1-etal}

\preprint{}

\title{Directionally Accelerated Detection of an Unknown Second Reactor with Antineutrinos for Mid-Field Nonproliferation Monitoring}


\author{D.~L.~Danielson}
\email[To whom correspondence should be addressed.\\E-mail: ]{daine@uchicago.edu}
\affiliation{Rare Event Detection Group, Lawrence Livermore National Laboratory, Livermore, California 94550, USA}
\affiliation{Department of Physics, University of California at Davis, Davis, California 95616, USA}
\affiliation{Theoretical Division, Los Alamos National Laboratory, Los Alamos, New Mexico 87545, USA}
\author{O.~A.~Akindele}\affiliation{Rare Event Detection Group, Lawrence Livermore National Laboratory, Livermore, California 94550, USA}
\author{M.~Askins}\affiliation{University of California at Berkeley, Berkeley, California 94720, USA}\affiliation{Lawrence Berkeley National Laboratory, Berkeley, California 94720, USA}\affiliation{Department of Physics, University of California at Davis, Davis, California 95616, USA}
\author{M.~Bergevin}\affiliation{Rare Event Detection Group, Lawrence Livermore National Laboratory, Livermore, California 94550, USA}
\author{A.~Bernstein}\affiliation{Rare Event Detection Group, Lawrence Livermore National Laboratory, Livermore, California 94550, USA}
\author{J.~Burns}\affiliation{Atomic Weapons Establishment, Aldermaston, Reading RG7 4PR, UK}
\author{A.~Carroll}\affiliation{University of Liverpool, Liverpool L69 3BX, UK}
\author{J.~Coleman}\affiliation{University of Liverpool, Liverpool L69 3BX, UK}
\author{R.~Collins}\affiliation{University of Liverpool, Liverpool L69 3BX, UK}
\author{C.~Connor}\affiliation{Boston University, Boston, Massachusetts  02215, USA}
\author{D.~F.~Cowen}
\affiliation{Department of Physics, Pennsylvania State University, University Park, Pennsylvania 16802, USA}
\affiliation{Department of Astronomy and Astrophysics, Pennsylvania State University, University Park, Pennsylvania 16802, USA}
\author{F.~Dalnoki-Veress}\affiliation{Middlebury Institute of International Studies at Monterey, Monterey, California 93940, USA}\affiliation{James Martin Center for Nonproliferation Studies, Monterey, California 93940, USA}
\author{S.~Dazeley}\affiliation{Rare Event Detection Group, Lawrence Livermore National Laboratory, Livermore, California 94550, USA}
\author{M.~V.~Diwan}\affiliation{Brookhaven National Laboratory, Upton, New York 11973, USA}
\author{J.~Duron}\affiliation{University of Hawai\okina i at M\={a}noa, Honolulu, Hawai\okina i 96822, USA}
\author{S.~T.~Dye}\affiliation{University of Hawai\okina i at M\={a}noa, Honolulu, Hawai\okina i 96822, USA}
\author{J.~Eisch}\affiliation{Iowa State University, Ames, Iowa 50011, USA}
\author{A.~Ezeribe}\affiliation{The University of Sheffield, Sheffield S10 2TN, UK}
\author{V.~Fischer}\affiliation{Department of Physics, University of California at Davis, Davis, California 95616, USA}
\author{R.~Foster}\affiliation{The University of Sheffield, Sheffield S10 2TN, UK}
\author{K.~Frankiewicz}\affiliation{Boston University, Boston, Massachusetts  02215, USA}
\author{C.~Grant}\affiliation{Boston University, Boston, Massachusetts  02215, USA}
\author{J.~Gribble}\affiliation{Atomic Weapons Establishment, Aldermaston, Reading RG7 4PR, UK}
\author{J.~He}\affiliation{Department of Physics, University of California at Davis, Davis, California 95616, USA}
\author{C.~Holligan}\affiliation{The University of Sheffield, Sheffield S10 2TN, UK}
\author{G.~Holt}\affiliation{University of Liverpool, Liverpool L69 3BX, UK}
\author{W.~Huang}\affiliation{University of Pennsylvania, Philadelphia, Pennsylvania 19104, USA}
\author{I.~Jovanovic}\affiliation{University of Michigan, Ann Arbor, Michigan 48109, USA}
\author{L.~Kneale}\affiliation{The University of Sheffield, Sheffield S10 2TN, UK}
\author{L.~Korkeila}\affiliation{Department of Physics, University of California at Davis, Davis, California 95616, USA}
\author{V.~A.~Kudryavtsev}\affiliation{The University of Sheffield, Sheffield S10 2TN, UK}
\author{P.~Kunkle}\affiliation{Boston University, Boston, Massachusetts  02215, USA}
\author{R.~Lap~Keung~Mak}\affiliation{Department of Physics, University of California at Davis, Davis, California 95616, USA}
\author{J.~G.~Learned}\affiliation{University of Hawai\okina i at M\={a}noa, Honolulu, Hawai\okina i 96822, USA}
\author{P.~Lewis}\affiliation{Atomic Weapons Establishment, Aldermaston, Reading RG7 4PR, UK}
\author{V.~A.~Li}\affiliation{Rare Event Detection Group, Lawrence Livermore National Laboratory, Livermore, California 94550, USA}
\author{X.~Liu}\affiliation{The University of Edinburgh, Edinburgh EH8 9YL, UK}
\author{M.~Malek}\affiliation{The University of Sheffield, Sheffield S10 2TN, UK}
\author{J.~Maricic}\affiliation{University of Hawai\okina i at M\={a}noa, Honolulu, Hawai\okina i 96822, USA}
\author{C.~Mauger}\affiliation{University of Pennsylvania, Philadelphia, Pennsylvania 19104, USA}
\author{N.~McCauley}\affiliation{University of Liverpool, Liverpool L69 3BX, UK}
\author{C.~Metelko}\affiliation{University of Liverpool, Liverpool L69 3BX, UK}
\author{R.~Mills}\affiliation{University of Liverpool, Liverpool L69 3BX, UK}
\author{F.~Muheim}\affiliation{The University of Edinburgh, Edinburgh EH8 9YL, UK}
\author{A.~St.\,J~Murphy}\affiliation{The University of Edinburgh, Edinburgh EH8 9YL, UK}
\author{M.~Needham}\affiliation{The University of Edinburgh, Edinburgh EH8 9YL, UK}
\author{K.~Nishimura}\affiliation{University of Hawai\okina i at M\={a}noa, Honolulu, Hawai\okina i 96822, USA}
\author{G.~D.~Orebi Gann}\affiliation{University of California at Berkeley, Berkeley, California 94720, USA}\affiliation{Lawrence Berkeley National Laboratory, Berkeley, California 94720, USA}
\author{S.~M.~Paling}\affiliation{Boulby Underground Laboratory, Loftus, Saltburn-by-the-Sea, Cleveland TS13 4UZ, UK}
\author{T.~Pershing}\affiliation{Department of Physics, University of California at Davis, Davis, California 95616, USA}
\author{L.~Pickard}\affiliation{Department of Physics, University of California at Davis, Davis, California 95616, USA}
\author{B.~Pinheiro}\affiliation{University of Pennsylvania, Philadelphia, Pennsylvania 19104, USA}
\author{S.~Quillin}\affiliation{Atomic Weapons Establishment, Aldermaston, Reading RG7 4PR, UK}
\author{S.~Rogers}\affiliation{University of Pennsylvania, Philadelphia, Pennsylvania 19104, USA}
\author{A.~Scarff}\affiliation{The University of Sheffield, Sheffield S10 2TN, UK}
\author{Y.~Schnellbach}\affiliation{University of Liverpool, Liverpool L69 3BX, UK}
\author{P.~R.~Scovell}\affiliation{Boulby Underground Laboratory, Loftus, Saltburn-by-the-Sea, Cleveland TS13 4UZ, UK}
\author{T.~Shaw}\affiliation{Atomic Weapons Establishment, Aldermaston, Reading RG7 4PR, UK}
\author{V.~Shebalin}\affiliation{University of Hawai\okina i at M\={a}noa, Honolulu, Hawai\okina i 96822, USA}
\author{G.~D.~Smith}\affiliation{The University of Edinburgh, Edinburgh EH8 9YL, UK}
\author{M.~B.~Smy}\affiliation{University of California at Irvine, Irvine, California 92697, USA}
\author{N.~Spooner}\affiliation{The University of Sheffield, Sheffield S10 2TN, UK}
\author{F.~Sutanto}\affiliation{University of Michigan, Ann Arbor, Michigan 48109, USA}
\author{R.~Svoboda}\affiliation{Department of Physics, University of California at Davis, Davis, California 95616, USA}
\author{L.~F.~Thompson}\affiliation{The University of Sheffield, Sheffield S10 2TN, UK}
\author{C.~Toth}\affiliation{Boulby Underground Laboratory, Loftus, Saltburn-by-the-Sea, Cleveland TS13 4UZ, UK}
\author{M.~R.~Vagins}\affiliation{University of California at Irvine, Irvine, California 92697, USA}
\author{S.~Ventura}\affiliation{University of Hawai\okina i at M\={a}noa, Honolulu, Hawai\okina i 96822, USA}
\author{M.~J.~Wetstein}\affiliation{Iowa State University, Ames, Iowa 50011, USA}
\author{M.~Yeh}\affiliation{Brookhaven National Laboratory, Upton, New York 11973, USA}

\collaboration{AIT-WATCHMAN Collaboration}
\noaffiliation

\date{\today}
\begin{abstract}
\noindent
When monitoring a reactor site for nuclear nonproliferation purposes, the presence of an unknown or hidden nuclear reactor could be obscured by the activities of a known reactor of much greater power nearby. Thus when monitoring reactor activities by the observation of antineutrino emissions, one must discriminate known background reactor fluxes from possible unknown reactor signals under investigation. To quantify this discrimination, we find the confidence to reject the (null) hypothesis of a single proximal reactor, by exploiting directional antineutrino signals in the presence of a second, unknown reactor. In particular, we simulate the inverse beta decay (IBD) response of a detector filled with a 1~kT fiducial mass of Gadolinium-doped liquid scintillator in mineral oil. We base the detector geometry on that of WATCHMAN, an upcoming antineutrino monitoring experiment soon to be deployed at the Boulby mine in the United Kingdom whose design and deployment will be detailed in a forthcoming white paper. From this simulation, we construct an analytical model of the IBD event distribution for the case of one $4\mathrm{\ GWt}\pm2\%$ reactor 25 km away from the detector site, and for an additional, unknown, 35~MWt reactor 3~to~5~km away. The effects of natural-background rejection cuts are approximated. Applying the model, we predict $3\sigma$ confidence to detect the presence of an unknown reactor within five weeks, at standoffs of 3~km or nearer. For more distant unknown reactors, the $3\sigma$ detection time increases significantly. However, the relative significance of directional sensitivity also increases, providing up to an eight week speedup to detect an unknown reactor at 5~km away. Therefore, directionally sensitive antineutrino monitoring can accelerate the mid-field detection of unknown reactors whose operation might otherwise be masked by more powerful reactors in the vicinity.
\end{abstract}



\maketitle
\section{Introduction}
In any antineutrino reactor-monitoring scenario, observers contend with low event rates, time constraints on observation, and backgrounds both artificial and natural. In particular, known energy-producing reactors can overwhelm signals of interest that might otherwise be observable to an antineutrino monitor. Thus, a nonproliferation monitoring program must characterize the requirements to detect the presence of a significant unknown reactor even in the vicinity of a known reactor (Figure 1).

	\begin{figure}
		\includegraphics[width=\linewidth]{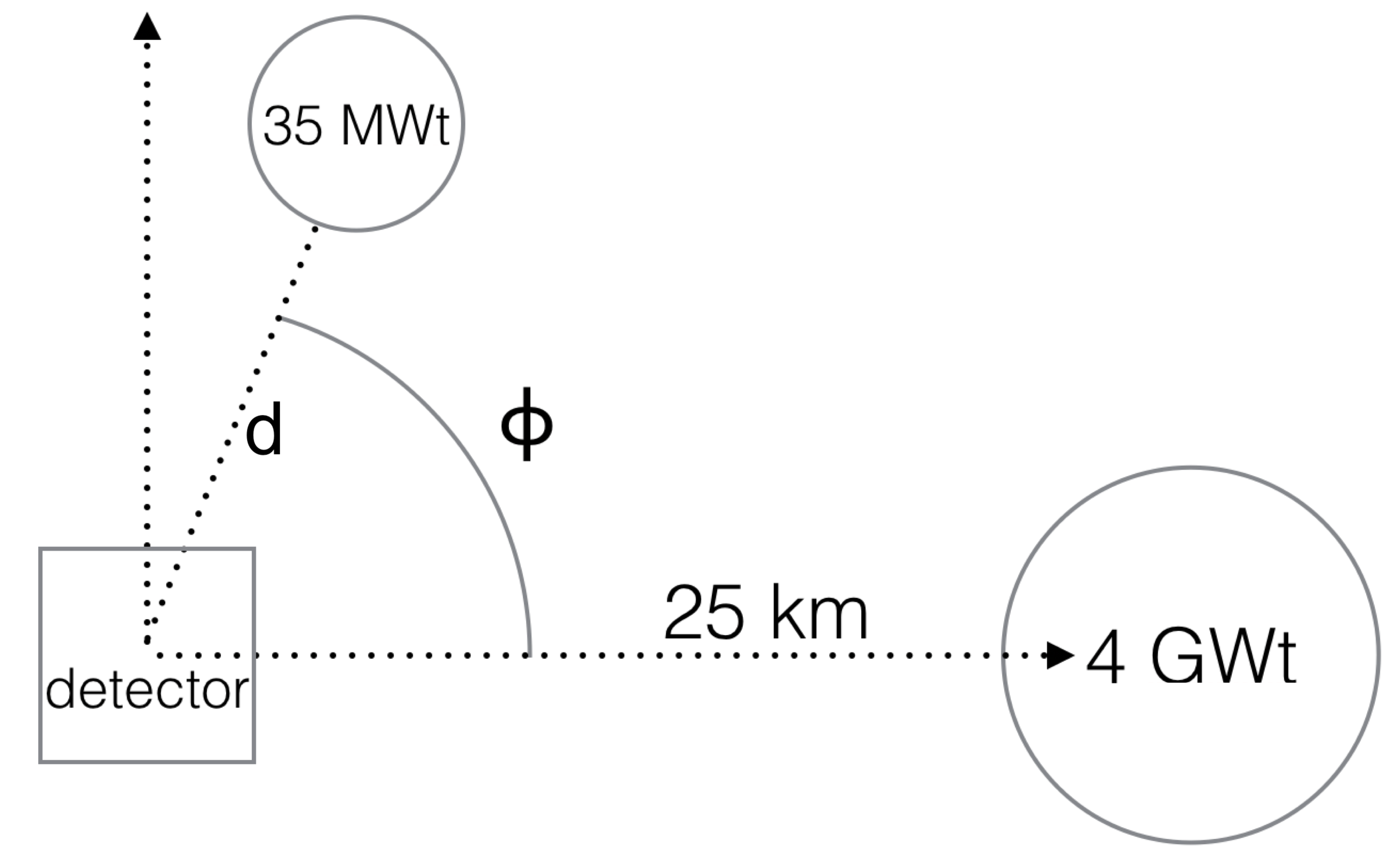}%
		\caption{The mid-field reactor configuration under study, relative to an antineutrino detector at the origin. A known 4~GWt reactor sits 25~km in the $x$-direction, with a second, unknown 35~MWt reactor at a standoff $d$, with an azimuthal separation $\phi$. We investigate the sensitivity of the detector to signify the presence of this second, unknown reactor.}
		
	\end{figure}

The WATCHMAN (Water Cherenkov Monitor for Anti-Neutrinos) collaboration exists to demonstrate the utility of water Cherenkov antineutrino monitoring for nuclear nonproliferation reactor monitoring \cite{watchman-1, watchman-2}. As part of this collaboration's broader interest in demonstrating the utility of antineutrino monitoring for nuclear nonproliferation, we investigate a detector with equivalent geometry to the WATCHMAN detector, but with an alternate fiducial composition.

To incorporate both event-rate and directional sensitivity, we consider a 1~kT-fiducial detector filled with Gadolinium-doped liquid scintillator dissolved in mineral oil. For antineutrinos of energy less than 15 MeV, we exploit the known weak preference for backwards positron emission \cite{PhysRevD.60.053003}, combined with the neutron's forward kinematics, to reconstruct a vector from the reconstructed neutron capture vertex to the centroid of positron scintillation. In the case of a single antineutrino source, the expectation direction of this vector points back towards the source reactor.

The purpose of this work is to characterize the sensitivity of such a detector to resolve the presented monitoring scenario using the direction-sensitive hypothesis testing techniques presented below, and to develop a simple analytical method for modeling IBD event spatial distributions in the development of future detectors and applications.

\section{Methods}
%
%
%
\subsection{Simulation and Modeling}
	\begin{figure}
		\includegraphics[width=\linewidth]{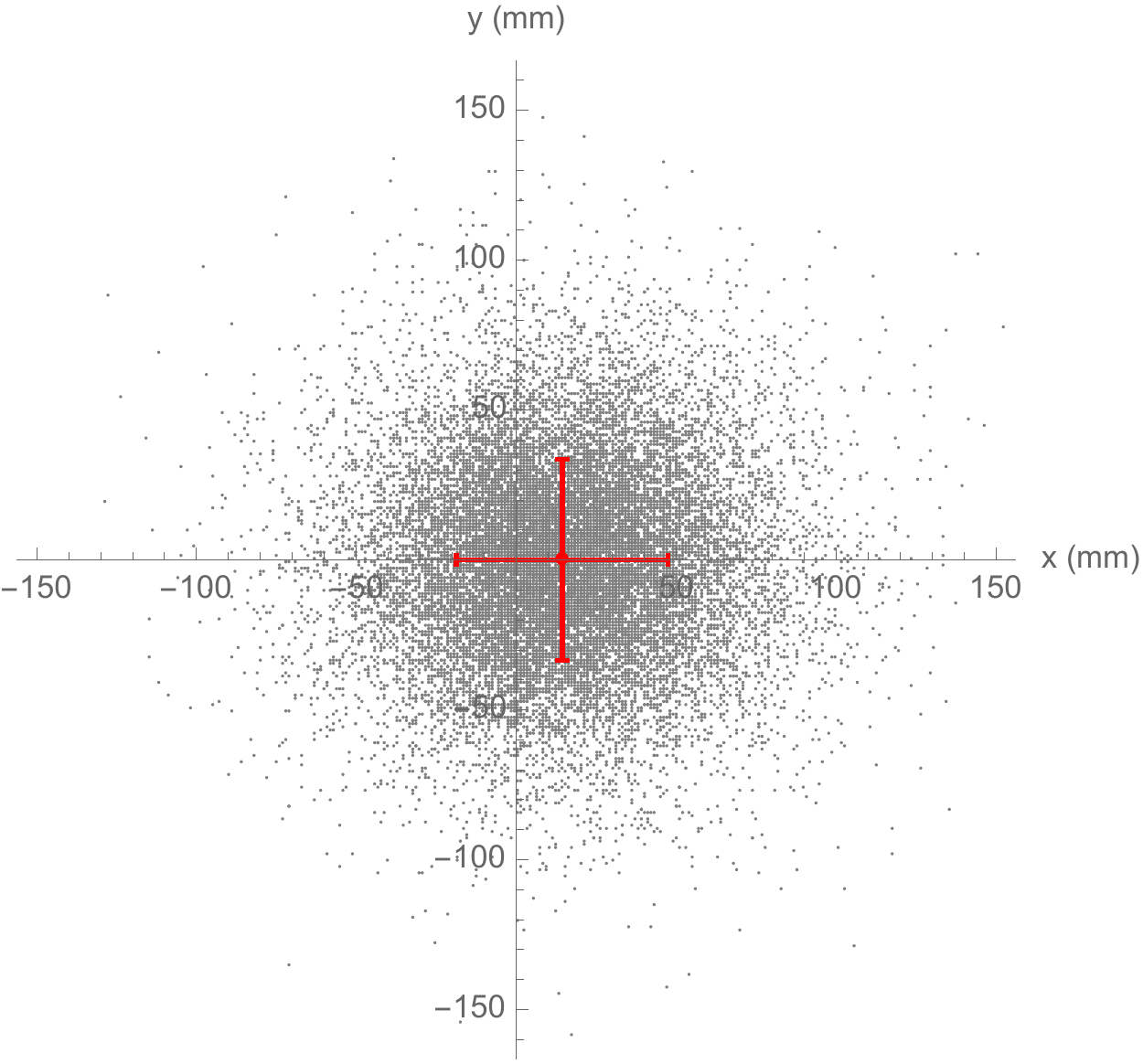}
		\caption{Relative positron-centroid $xy$-positions for 20,000 simulated inverse beta decays from reactor-spectrum~\cite{RevModPhys.74.297} antineutrinos. Events take place in a 16~m~$\times$~16~m cylindrical volume of mineral oil, doped with 0.100\%~Gadolinium and 0.177\%~PPO~scintillator dissolved in 19.8\%~pseudocumene. We plot the centroid of scintillation for each positron, relative to its coincident neutron capture--defining the origin. Red error bars indicate the population standard deviation, centered on the population mean.}
	\end{figure}

	We consider an antineutrino detector sensitive to the positron \& neutron-capture double coincidence characteristic of inverse beta decay (IBD). In particular, we model the response of such a detector to multiple fission reactors, at arbitrary positions. Applying this model, we will analyze the expected response of our detector in the presence of two reactors given azimuthal separation $\phi$, and each $i^\mathrm{th}$ reactor's detected integrated flux $N_i$.

	First we consider the single-reactor case, from which to construct the $n$-reactor case by linear superposition. Using Geant4 \cite{AGOSTINELLI2003250} via rat-pac \cite{Seibert:2018aa}, we simulate the positron tracks and positron centroid vertices resulting from reactor-spectrum~\cite{RevModPhys.74.297} antineutrinos interacting in a 16~m~$\times$~16~m cylindrical volume of mineral oil, doped with 0.100\%~Gadolinium by mass and 0.177\%~PPO~scintillator dissolved in 19.8\%~pseudocumene. Figure~2 shows the resulting $xy$-projected positron track midpoints, plotted relative to the respective neutron capture vertices--defining the origin. We model the distribution of these vertices by fitting an uncorrelated tri-gaussian to them, thus obtaining the distribution parameters given in Table~I. Hereafter, we call this the ``positron centroid distribution."
	
			By modeling the positron centroid distribution as a single tri-gaussian, we implicitly `average' over some higher-order corrections due to the energy dependance of the neutron thermalization length \cite{PhysRevD.60.053003}. Similarly, by drawing neutron energies from a reactor IBD spectrum before neutrino oscillations, we neglect thermalization length modulations arising from those oscillations, these being higher-order corrections at the baselines under consideration ($\lesssim 0.1\%$ effect on time to $3\sigma$ detection). Importantly, however, we retain the primary effect of oscillations to modulate the IBD event rate.

		\begin{figure}
		\includegraphics[width=\linewidth]{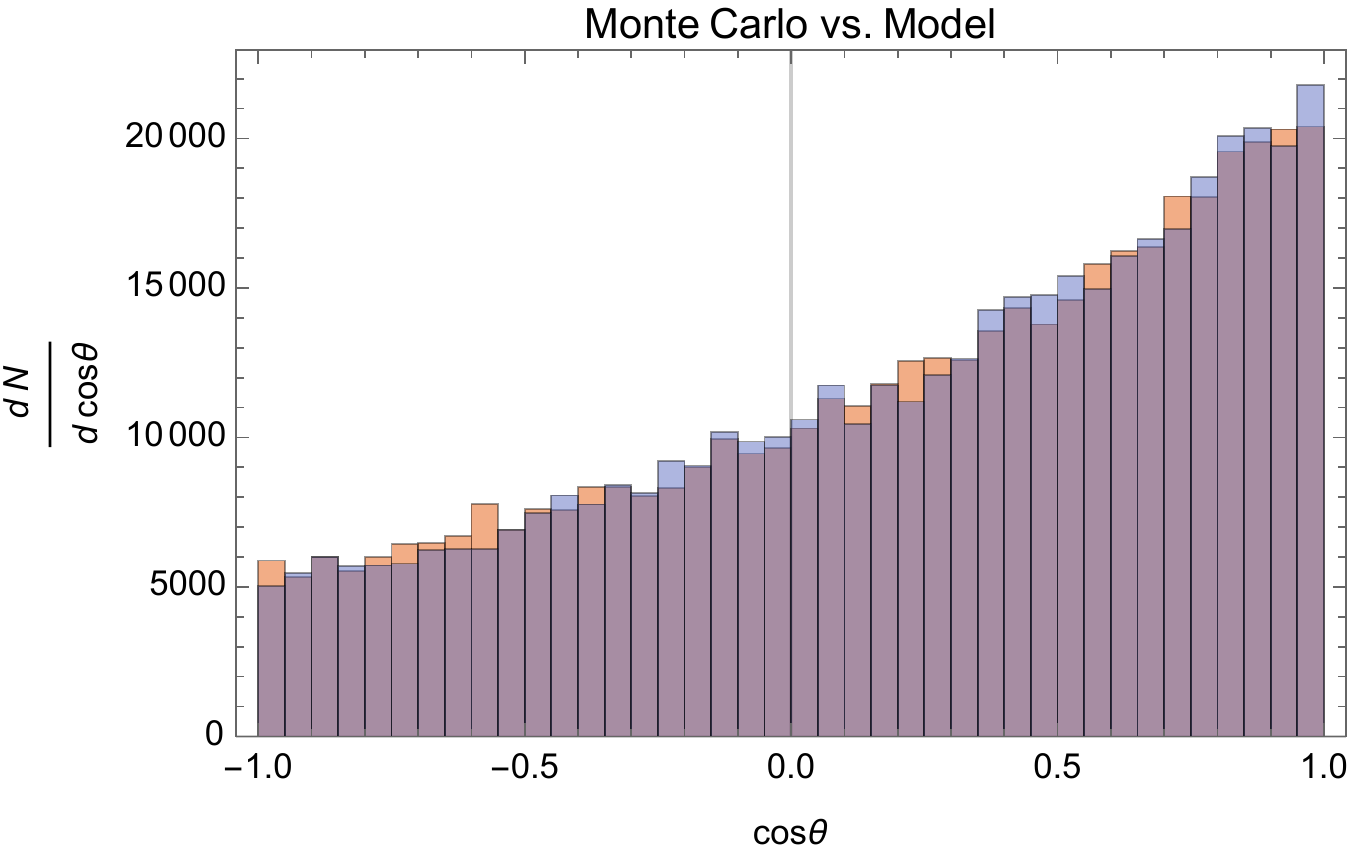}%
		\caption{Cosine density histograms for rat-pac/Geant4 Monte Carlo (blue) and samples drawn from the cosine marginal distribution of our model (orange). The good agreement ($\chi^2/\mathrm{dof}$=1) validates our analytical model~(1) against Monte Carlo truth prior to vertex reconstruction uncertainty~(3).}
		
	\end{figure}

	\begin{table}
		\caption{Uncorrelated tri-gaussian parameters describing positron midpoint positions relative to neutron captures.}
		\begin{ruledtabular}
			\begin{tabular}{l | r | r}
			 	     $q$ & $\mu_q$ (mm) & $\sigma_q$ (mm)\\
				\hline
	        	$x$ & $14.4 \pm0.2$ & $33.1\pm0.2$ \\
	        	$y$ & $0.1 \pm0.2$ & $33.5\pm0.2$ \\
		        $z$ & $-0.3 \pm0.2$ & $33.1\pm0.2$ \\
			\end{tabular}
		\end{ruledtabular}
	\end{table}
	
	Given the parameters in Table~I, and noting $\sigma_x\cong\sigma_y\cong\sigma_z\equiv\sigma$, we therefore model the single-reactor (n=1) positron centroid distribution by,
	
	\begin{equation}
		\frac{dN}{d^3r}\left(\vec{r}|n\!=\!1\right)=\frac{N}{\sqrt{2\pi\sigma^2}^3}e^{\frac{-(\vec{r}-\vec{\mu})^2}{2\sigma^2}}
	,\end{equation}                    
\noindent where $N$ is the detected integrated flux and $d^3r=dxdydz=r^2\sin \theta d\theta d\phi dr$.
	By extension, the $n$-reactor case samples a linear combination of distributions of the same form, given by,
	
	\begin{equation}
		\frac{dN}{d^3r}\left(\vec{r}\right|n)=\sum\limits_{i=1}^n\frac{N_i}{\sqrt{2\pi\sigma_i^2}^3}e^{\frac{-(\vec{r}-\vec{\mu}_i)^2}{2\sigma_i^2}}
	,\end{equation}	
where $\vec{\mu}_i$ and $\sigma_i$ now denote the mean and standard deviation, respectively, of the $i^\mathrm{th}$ reactor's positron centroid distribution, along any direction. The isotropic variance of the positron centroid distribution described in Table~I suggests the approximately reactor-independent magnitudes $\sigma_i\approx
	\sigma_j\equiv\sigma$ and $\mu_i\approx\mu_j\equiv\mu$, where $i$ and $j$ indicate different reactors.

	To model the effect of vertex position reconstruction error, we marginalize $dN/d^3r$ over an uncorrelated tri-gaussian with $\sqrt{\sum_i{{\delta q_i}^2}}=\delta r\sqrt{2}$ and $\delta q_i\approx\delta q_j=\delta r\sqrt{2/3}$ where $\delta r$ denotes the vertex reconstruction resolution, and the factor of $\sqrt 2$ arises from the geometric sum of uncorrelated errors on the two reconstructed vertices in an IBD double-coincidence. We neglect the anisotropy of the positron midpoint reconstruction error along its scintillation track, on the grounds that, for the antineutrino energies under consideration, the positron track length is negligibly small relative to the vertex resolution itself~\cite{Apollonio:1999jg}. For a reconstructed vertex $\tilde r$,
	
	\begin{multline}
		\frac{dN}{d^3\tilde{r}}\left(\tilde{r}|n\right)=\frac{1}{\sqrt{2\pi\left(\delta r \sqrt{2/3}\right)^2}^3}\int\limits_{\forall \vec r} e^\frac{-(\tilde{r}-\vec{r})^2}{2(\delta r \sqrt{2/3})^2}\frac{dN}{d^3r}\left(\vec{r}\right|n)d^3r 
		\\ = \sum\limits_{i=1}^n\frac{N_i e^\frac{-\left(\tilde{r}-\vec{\mu}_i\right)^2}{2\left[\sigma^2 + \left(\delta r \sqrt{2/3}\right)^2\right]}}{\sqrt{2\pi\left[\sigma^2+\left(\delta r\sqrt{2/3}\right)^2\right]}^3}
	\end{multline}
	
	Finally, since we have already shown the positron centroid displacement magnitudes $\mu_i\approx\mu_j\equiv\mu$ contain negligible reactor discriminating power, we marginalize over the radius $|\tilde r|$ and the symmetry angle $\varphi'$ to concern ourselves only with the reconstructed polar cosine distribution,
\nopagebreak
	\begin{equation}
		\frac{dN}{d\cos\tilde\theta}\left(\cos\tilde\theta|n\right) =
		\int\limits_0^\infty\int\limits_{0}^{2\pi}\frac{dN}{d^3\tilde{r}}\left(\tilde{r}\right|n)|\tilde r|^2 d\varphi' d|\tilde r|
		,\end{equation}

	Figure~3 justifies our analytical model by showing good agreement with Monte Carlo truth prior to vertex reconstruction uncertainty.

	 To scale each reactor's contribution to $N$, neglecting core evolution, we approximate each $i^\mathrm{th}$ reactor's mean detected IBD count $\left<N_i\right>$ after a duration $t$ as,
\nopagebreak
\begin{equation}
\left<N_i\right>(t) \approx \varepsilon n_\mathrm{p} \frac{t}{4\pi L_i^2} \frac{p_i}{\epsilon} \int_0^\infty \sigma(E) P_\mathrm{ee}(L_i/E)\varphi(E) dE
,\end{equation}

\noindent where $\varepsilon$ is the total IBD detection efficiency, $n_\mathrm{p}$ is the number of free fiducial protons, $L_i$ is the propagation baseline, $p_i$ is the reactor's thermal power, $\epsilon=200\mathrm{\ MeV}$ approximates the average fission energy, $E$ is antineutrino energy, $\sigma$ models the IBD cross section, $P_\mathrm{ee}$ is the electron antineutrino survival probability, and $\varphi$ models the fractional reactor flux density~\cite{RevModPhys.74.297}.

	We model the IBD cross section as given in reference~\cite{ibd},
	\begin{equation}
	\begin{split}
	k_1&=0.07056\\
	k_2&=0.02018\\
	k_3&=0.001953\\
	E_e&=E_\nu-(m_n-m_p)\\
	\sigma(\bar \nu_e p)&\approx10^{-43}\mathrm{cm}^2p_eE_eE_\nu^{-k_1+k_2\ln E_\nu-k_3\ln^3E_\nu}
	\end{split}
	,\end{equation}
	where all energies and masses are expressed in MeV. This model is accurate to within 0.5\% for reactor antineutrino energies \cite{ibd}, and we include this systematic uncertainty in our analysis and in Table~II.
	
	Without loss of generality, we restrict ourselves to the inverted mass hierarchy case. We find a negligible event rate discrepancy of less than $0.4\%$ between the hierarchies. However, for the baselines under consideration, the inverted hierarchy yields up to $5\%$ larger systematic uncertainties, so we choose it as the conservative case.
	
	For given detected integrated fluxes $N_i$, we calculate the expected histogram by integrating (4) into bins of width $\Delta\cos\tilde\theta$ as justified by Freedman and Diaconis \cite{Freedman1981},

	\begin{equation}
	\Delta\cos\tilde{\theta}=2\frac{\mathrm{IQR}}{\sqrt[3]{\sum_i N_i}},
	\end{equation}
	where IQR is the interquartile range of the reconstructed polar cosine distribution. We give this histogram both total flux (correlated) and Poisson (uncorrelated) bin uncertainties.
	
		\begin{figure} 
		\begin{subfigure}{.34\textwidth}
		\centering
		\includegraphics[width=\linewidth]{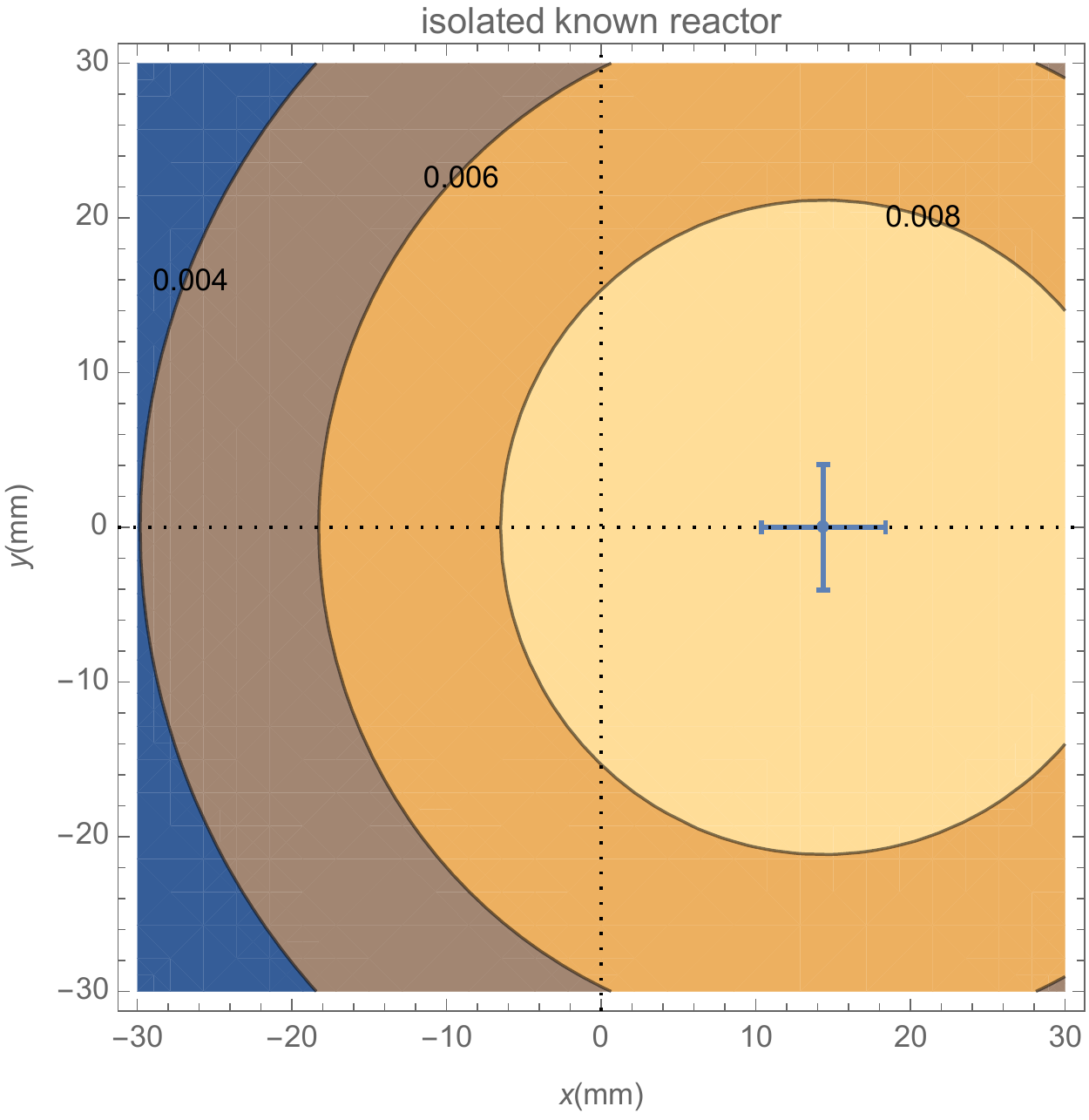}
		\caption{The null hypothesis: a single reactor signal in the positive $x$-direction.}
		\end{subfigure}
		\begin{subfigure}{.34\textwidth}
		\centering
		\includegraphics[width=\linewidth]{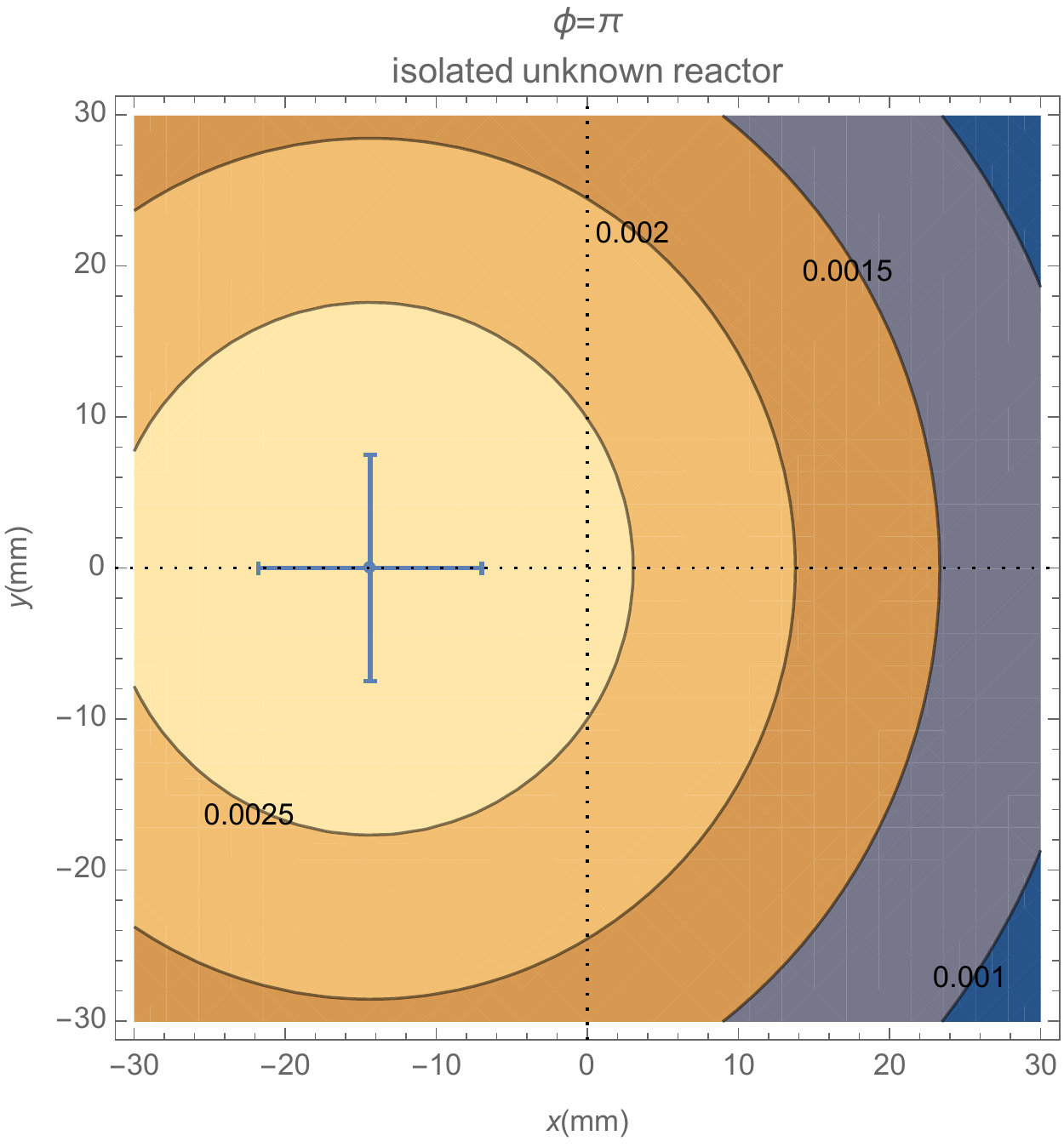}
	    \caption{The unknown reactor signal, in isolation, shown for a reactor in the negative $x$-direction~($\phi=\pi$).}
		\end{subfigure}
		\caption{IBD positron centroid event count $xy$-distributions for (a) the known $4\mathrm{\ GWt}\pm2\%$ reactor at a 25km standoff, and (b) the unknown 35 MWt reactor in isolation at a 5 km standoff, both after one month of exposure. The origin marks the neutron capture vertex, by definition. Error bars show the standard deviation on the mean of each distribution, centered on the mean. Contours values denote the constant value of the distribution along them. The color scale is reset between (a) and (b) to accommodate their different total event counts, as evidenced by the contour values.}
	\end{figure}

E. Caden's work on behalf of the Double Chooz collaboration demonstrates that judicious cuts can reduce natural background-induced effects to negligible levels in directionality studies \cite{caden}. Hence the effect of natural backgrounds is approximated through their impact on the overall detection efficiency $\varepsilon$.

\subsection{Hypothesis Testing}

	To characterize the sensitivity of our detector to distinguishing a two-reactor scenario from the one-reactor (null) hypothesis, we consider the one-sided 95\% confidence interval (CI) limit on significance to reject the null hypothesis, quantifying significance as the $Z$-score of the reduced chi-square test statistic~\cite{Olive},
	\begin{equation}
	Z=\Phi^{-1}\left[1-\int\limits_{\chi^2}^\infty f(z; \mathrm{dof})dz\right]
	,\end{equation}
\noindent where $\Phi^{-1}$ is the quantile of the standard Gaussian, and $f$ is the $\chi^2$ probability density function. To predict the one-sided confidence interval limit on $Z$ with a confidence level CL, we use a conservative approximation following~\cite{denker}. For an upper limit,
	\begin{equation}
	Z_{\mathrm{CL}} \lesssim Z \rvert_{N_i=\left<N_i\right> +\delta N_i\Phi^{-1}(1-\mathrm{CL})}
	,\end{equation}
where $Z_{\mathrm{CL}}$ is the limit, and from the $i^{\mathrm{th}}$ reactor $\left<N_i\right>$ counts the expected IBD detections with a total uncertainty $\delta N_i$. For a lower limit, we reverse the inequality and subtract $\delta N_i$ instead.
	
\subsubsection{Systematic Uncertainties}

\begin{table}
		\caption{Systematic Uncertainties in Expected IBD Event Count}
		\begin{ruledtabular}
			\begin{tabular}{r | l}
			 	     source & systematic uncertainty $(\sigma_\mathrm s)_i / \left<N\right>$ \\
				\hline
				thermal power & $2\%$ \\
				IBD cross section & $0.5\%$ \cite{ibd} \\
		        reactor neutrino anomaly & $2\%$  \\
		        oscillation parameters & See reference~\cite{CAPOZZI2016218}
			\end{tabular}
		\end{ruledtabular}
	\end{table}

Systematic uncertainties accumulate from thermal power IBD conversion, cross section uncertainties \cite{ibd}, flavor oscillation uncertainties, and any reactor-anomaly shift. Each of these modulates the expectation number of total IBD events in the single reactor (null) hypothesis~(5). We apply the systematic uncertainties shown in Table~II:
\begin{equation}
    N(t|n\!=\!1) \:=\: \left<N_1\right>(t) \:\pm\: \sqrt{\left<N_1\right>(t)} \:\pm\: \sqrt{\sum_m (\sigma_\mathrm s)^2_m}
,\end{equation}

\noindent where $\left<N_1\right>(t)$ counts the expected IBD detections due to the known reactor, after a period $t$. Since none of these systematics $(\sigma_\mathrm s)_m$ exhibit angular dependence, they manifest as bin-to-bin correlated uncertainties when histogramming the null hypothesis. To accommodate bin-to-bin correlated uncertainties, we apply the generalized form of $\chi^2$ when testing the null hypothesis,

\begin{equation}
\chi^2 \equiv (O_j-E_j)[\sigma_{kj}^2]^{-1}(O_k-E_k)
,\end{equation}
\nopagebreak
\noindent where $O_j$ is the observed histogram, $E_j$ is the null hypothesis expectation, $\sigma^2_{kj}$ is the covariance matrix~\cite{fogli}, and $j, k$ are bin indices under Einstein summation.
	
	\begin{figure}
	    \begin{subfigure}{.35\textwidth}
		\includegraphics[width=\linewidth]{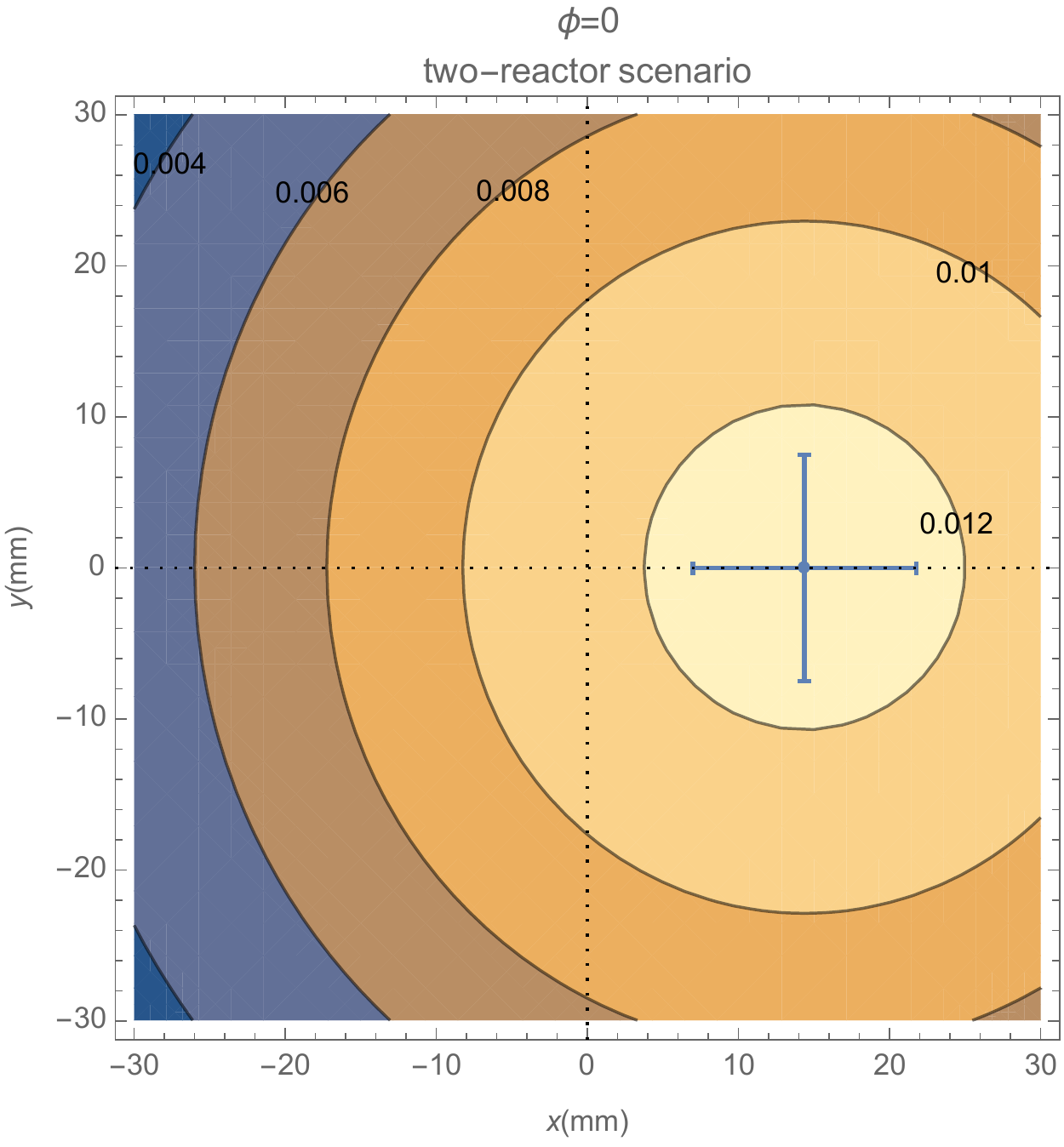}
		\end{subfigure}
	    \begin{subfigure}{.35\textwidth}
		\includegraphics[width=\linewidth]{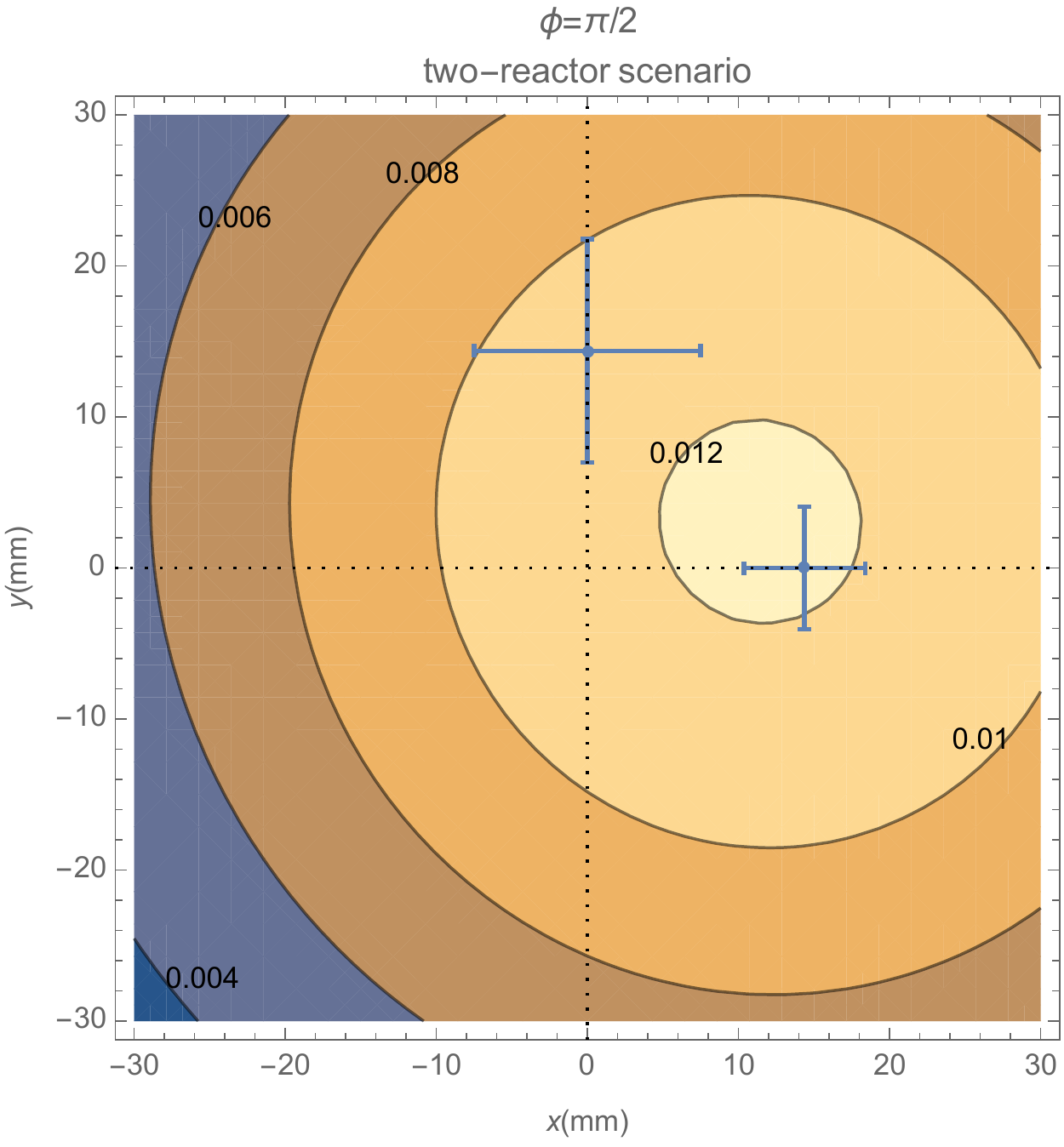}
		\end{subfigure}
	    \begin{subfigure}{.35\textwidth}
		\includegraphics[width=\linewidth]{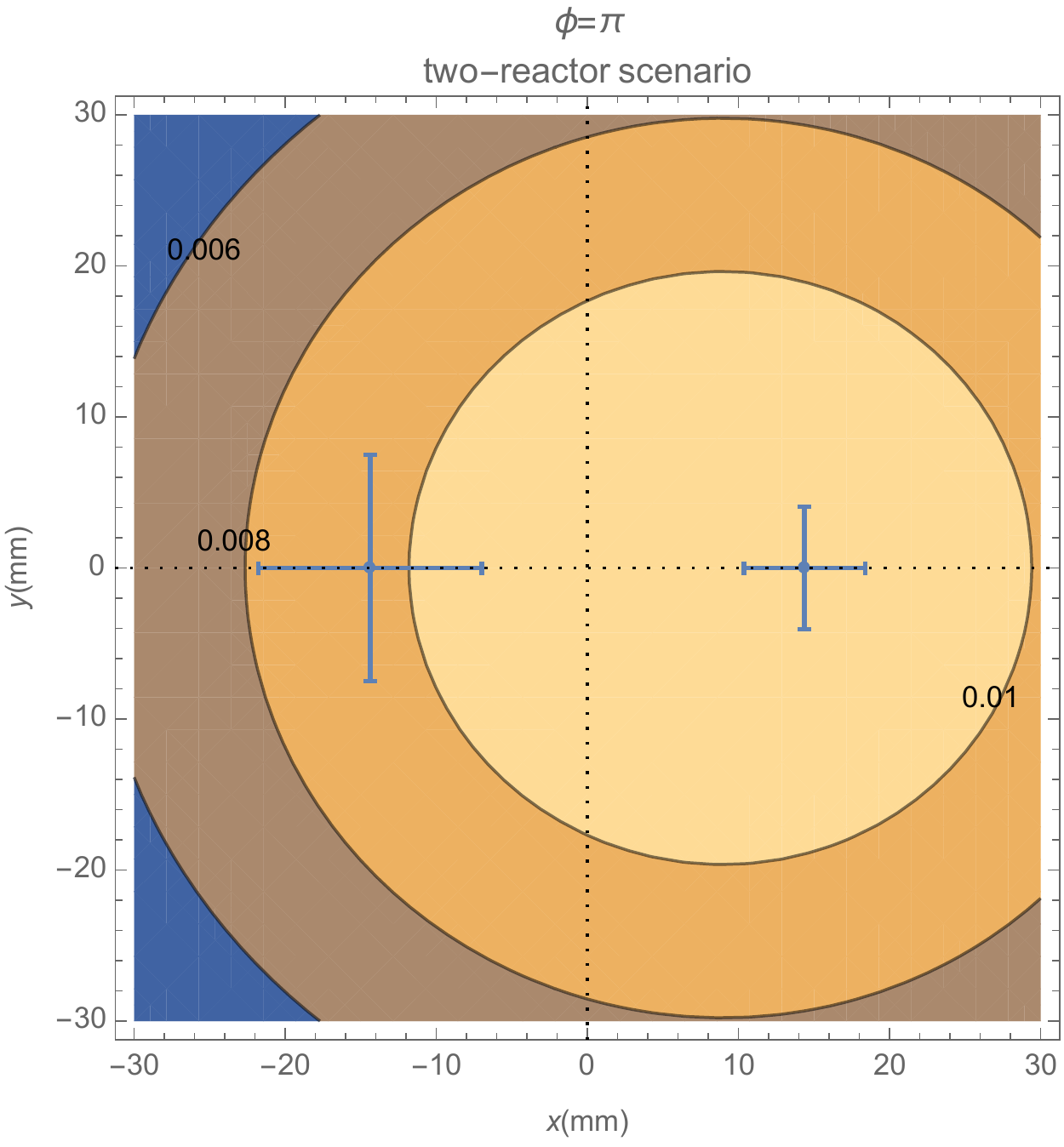}%
		\end{subfigure}
		\caption{IBD positron centroid event count $xy$-distributions for two-reactor scenarios, varying the azimuthal separation $\phi\in\{0,\pi/2,\pi\}$ between the known and unknown reactors. The known, 4 GWt reactor sits at a 25~km standoff, while the unknown 35 MWt reactor sits at a $d=5\mathrm{\ km}$ standoff, all after one month of exposure. We define the origin and error bars as in Figure~4.}
	\end{figure}
	Figure~5 presents two-reactor scenarios, showing the superposition of these distributions for a selection of azimuthal separations $\phi$ and a fixed unknown reactor standoff $d=5\mathrm{\ km}$.

\section{Results}

	For this scenario we assume a total IBD detection efficiency of $\varepsilon=80\%$ after detector losses and background rejection cuts, and a Daya Bay~\cite{Littlejohn:2012aa} or Double Chooz~\cite{Gomez}-like spatial resolution of $\delta r=15 \mathrm{\ cm}$. This justifies our previous assertion that the positron reconstruction error is only negligibly anisotropic: the entire positron scintillation track is $\sim0.5\mathrm{\ mm}$ for reactor antineutrino energies~\cite{Apollonio:1999jg}, which is negligible relative to this 15~cm vertex resolution. When calculating the propagation baseline for each standoff, we assume a 500 m detector overburden.
	
	Figure~4~(a) shows the single-reactor positron centroid $xy$-distribution after one month of exposure to a 4~GWt fission reactor located 25~km in the positive $x$-direction, and (b) shows the same for exposure to a 35~MWt fission reactor located 5~km in the negative $x$-direction. Given the high-power reactor's presence is known, Figure~4~(a) depicts the null hypothesis for detection of the additional low-power reactor's presence.

	\begin{figure}
		\begin{subfigure}{.4\textwidth}
		\includegraphics[width=\linewidth]{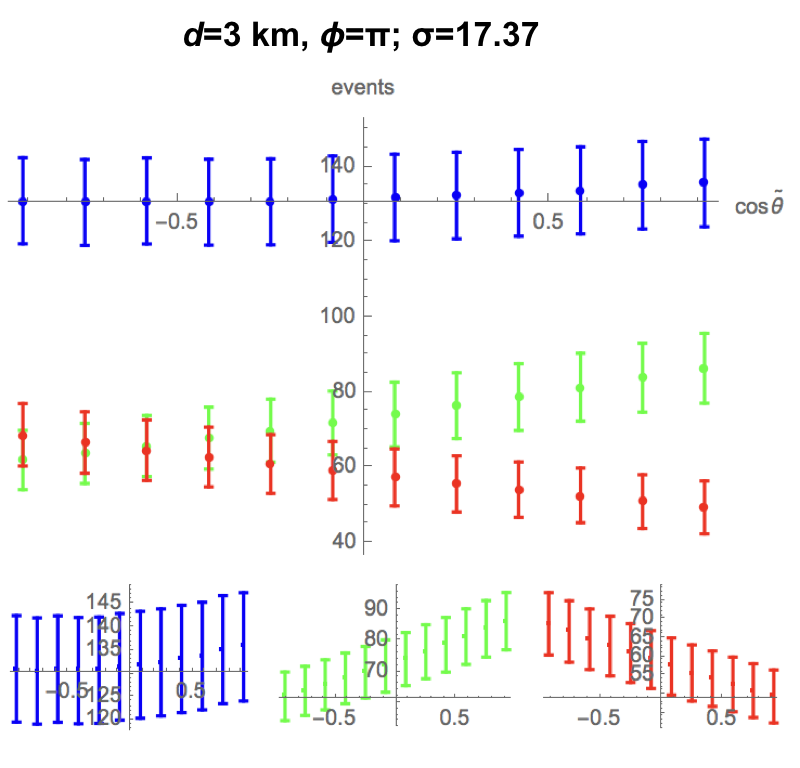}
		\end{subfigure}
		\begin{subfigure}{.4\textwidth}
		\includegraphics[width=\linewidth]{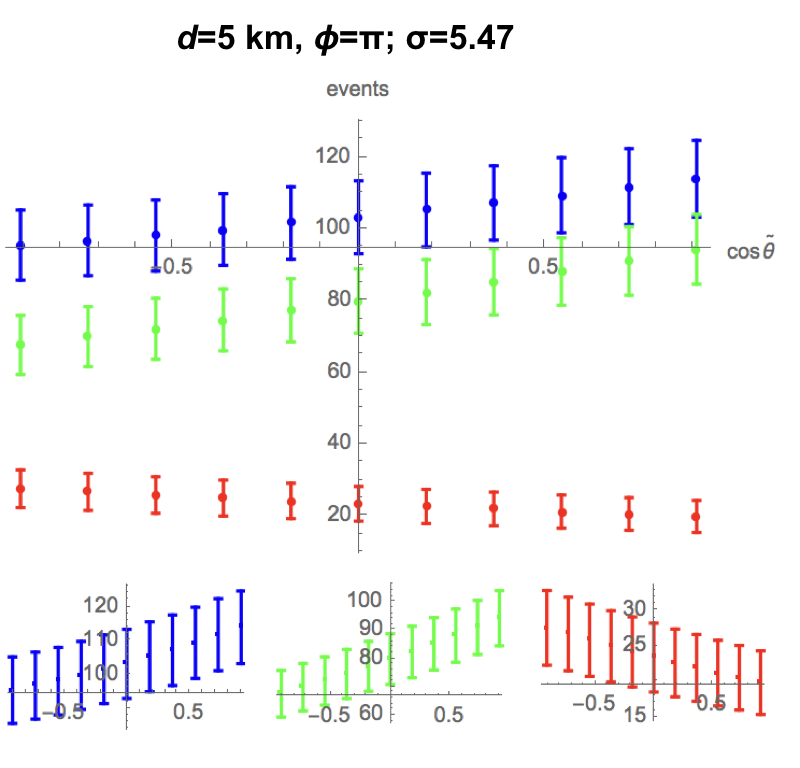}
		\end{subfigure}
		\caption{Predicted reconstructed cosine histograms after one year monitoring two reactors with an azimuthal separation $\phi=\pi$, for unknown-reactor standoffs $d=3\mathrm{\ km}$ (top) and $d=5\mathrm{\ km}$ (bottom). Larger $\sigma$ values indicate a stronger significance to detect the unknown reactor (reject the null hypothesis).\\We test the two-reactor total signal (blue, left insets for detail) against the single reactor (null) hypothesis (green, center insets for detail). For completeness, we also plot the unknown-reactor signal component (red, right insets for detail). We plot only statistical uncertainties, whereas $\sigma$ accounts for both systematic and statistical uncertainties. See Appendix A for a range of azimuthal separations $\phi\in\{0,\pi/4,\pi/2,3\pi/4\}$.}
	\end{figure}
	
	Figure~6 presents expected two-reactor and one-reactor reconstructed cosine histograms after one year of monitoring, for the maximum azimuthal separation~($\phi=\pi$), at unknown reactor standoffs of $d=3\mathrm{\ km}$ and $d=5\mathrm{\ km}$.

\section{Discussion}
	\begin{figure}
			\includegraphics[width=\linewidth]{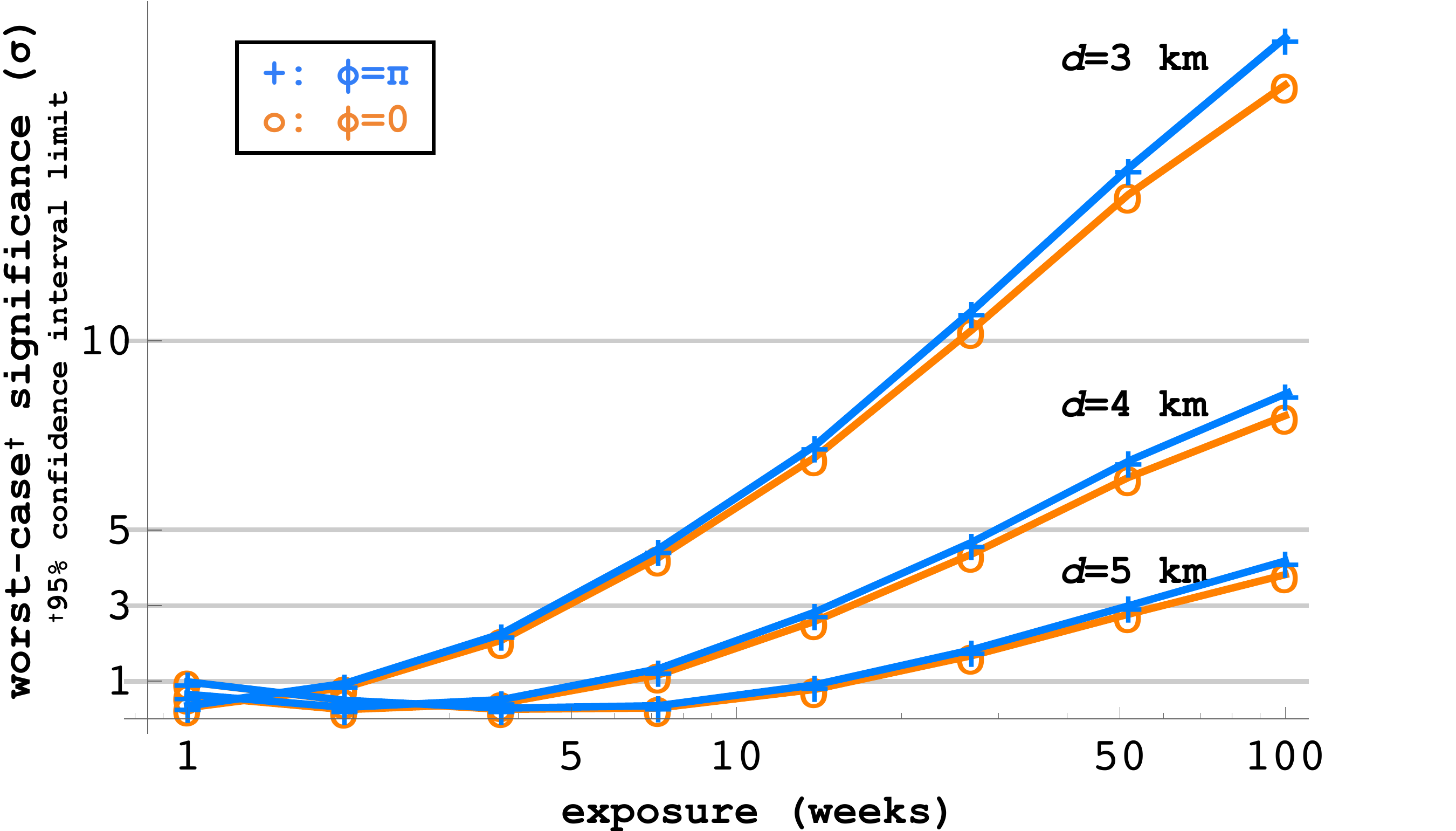}%
			\caption{Rejection of the single-reactor (null) hypothesis over time in the presence of an unknown 35 MWt reactor at $d \in \{3,4,5\}$ km for $\phi \in \{0,\pi\}$, given a known 4 GWt reactor at 25 km. We plot the lower limit of the 95\% confidence interval, as integrated downward from $\sigma=+\infty$, thus giving a worst-case statistical bound.}
	\end{figure}
		
	For the detector under consideration, the time to achieve $3\sigma$ detection of the unknown reactor depends more strongly on the detector-reactor standoff than it does on directional discrimination. Figure~7 shows that for a reactor-detector standoff of 3~km, $3\sigma$ detection is likely (95\% CL) within 5 weeks. However, the time required extends through 15 weeks ($\phi = \pi$) to 16 weeks ($\phi = 0$) for 4~km, and 52~weeks ($\phi=\pi$) to 60~weeks ($\phi=0$) for 5~km. When compared to this strong standoff dependence, directional discrimination plays a negligible role at 3~km, increasing towards an 8 week speedup at 5~km. We must note, however, that a detector design more optimized for direction reconstruction could provide a greater improvement in significance for $\phi > 0$.
	
	\begin{figure}
		\includegraphics[width=\linewidth]{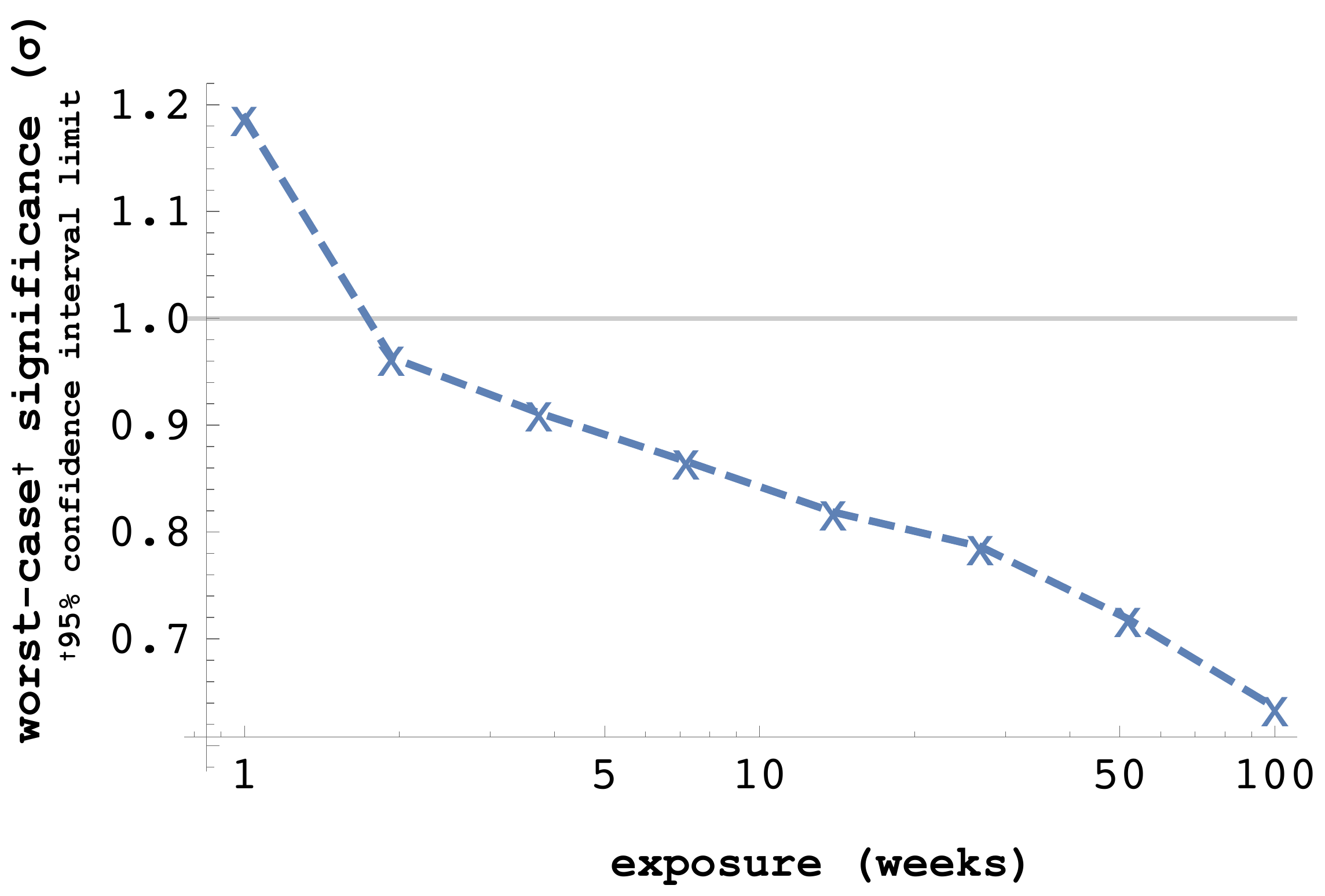}%
		\caption{Demonstration that our method carries no bias towards detection, by its failure to reject the null hypothesis over time in the absence of a second reactor, for a single known 4~GWt reactor at 25~km. We plot the upper limit of the 95\% confidence interval from $-\infty$.}
			\end{figure}

	Finally, Figure~8 shows the failure to reject the null hypothesis in the absence of a second reactor. In this case, in which the null hypothesis is in fact true, the confidence interval limit begins at $\sim1\sigma$ and decreases monotonically. This demonstrates that our method carries no bias towards false-positive detection.
	
\section{Outlook}
\subsection{Techniques}
	We have simulated a WATCHMAN-based detector geometry containing Gadolinium-doped liquid scintillator, and characterized its spatial IBD response to an isotropic antineutrino source. From this we have derived a statistical model describing the true detector event positions, their reconstructed vertices, and the reconstructed directionality. This statistical analysis accepts, as input parameters: the number and positions of source reactors, the true-position distributions, the achievable vertex resolution, the overall detector efficiency, and arbitrary additional correlated and uncorrelated uncertainties. We have developed a parallel-computing Mathematica \cite{Mathematica} code to ease the efficient production of further analyses based on this statistical model. 
	
	In future, we see natural applications for this model in running similar analyses on various reactor-detector configurations.
\subsection{Nonproliferation}
	We have characterized the sensitivity of a 1~kT fiducial volume of GdLS mineral oil to detect the presence of a second, unknown reactor, for a range of configurations on the ground. We have shown evidence for feasible month-scale rapid detection of an unknown, second reactor in the presence of a more powerful, known reactor, provided sufficient proximity ($d \le 3 \mathrm{\ km}$) to the unknown source. For more distant standoffs, the time to detect increases significantly. We have also shown that, for our detector design, the directional sensitivity inherent to our chosen fiducial material contributes only marginally in discovering an unknown reactor azimuthally separated from a known one at 3~km, but that increasing the unknown standoff through 5 km increases the relative contribution of directional sensitivity, providing up to a 24\% speedup for $\phi=\pi$ relative to $\phi=0$ (Figure~7).
	
	Looking beyond this work, it is reasonable to expect that more directionally-optimized detector configurations may achieve the order-of-magnitude reduction in detection time required to enable month-scale monitoring for the presence of unknown reactors at farther mid-field standoffs. Future studies by the WATCHMAN collaboration will explore multi-detector and segmented-detector deployments to characterize their effectiveness at this task.
\nopagebreak
\begin{acknowledgments}
D.L.D. thanks the Rare Event Detection Group at Lawrence Livermore National Laboratory for their patience while he builds his company.
He also acknowledges his support as a Nuclear Science and Security Consortium Fellow.

We also thank the U.K.'s National Nuclear Laboratory for support of Robert Mills.

This material is based upon work supported by the U.S. Department of Energy National Nuclear Security Administration under Award Number DE-NA0000979 through the Nuclear Science and Security Consortium.
This work was performed under the auspices of the U.S. Department of Energy by Lawrence Livermore National Laboratory under Contract DE-AC52-07NA27344.
\end{acknowledgments}
\vfill
\FloatBarrier
\appendix
\clearpage
\section{\vspace{-10pt}}
	\begin{figure}[h!]
	    \begin{subfigure}{.27\textwidth}
		\includegraphics[width=\linewidth]{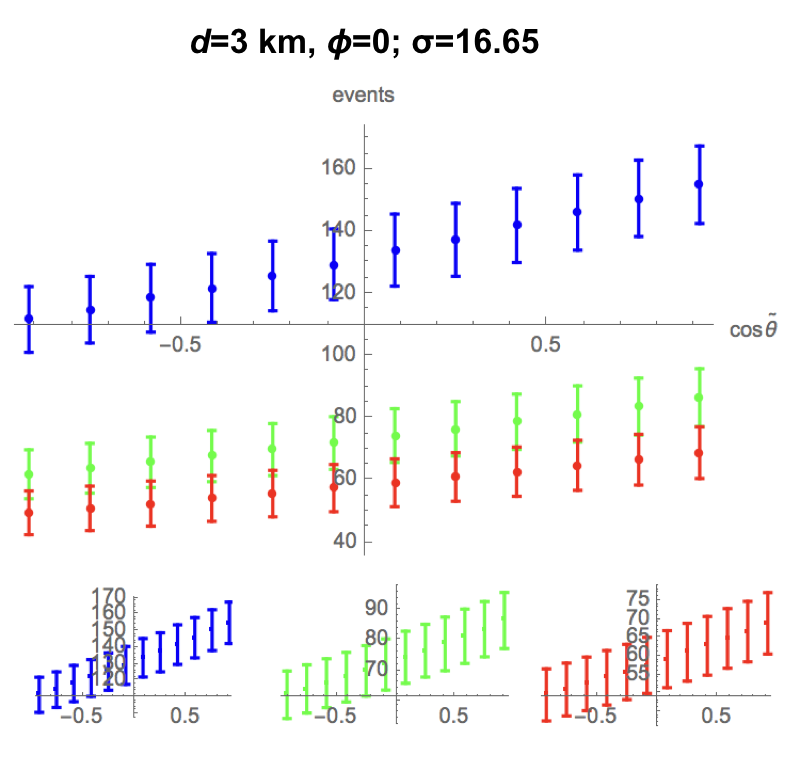}
		\end{subfigure}\\
	    \begin{subfigure}{.27\textwidth}
		\includegraphics[width=\linewidth]{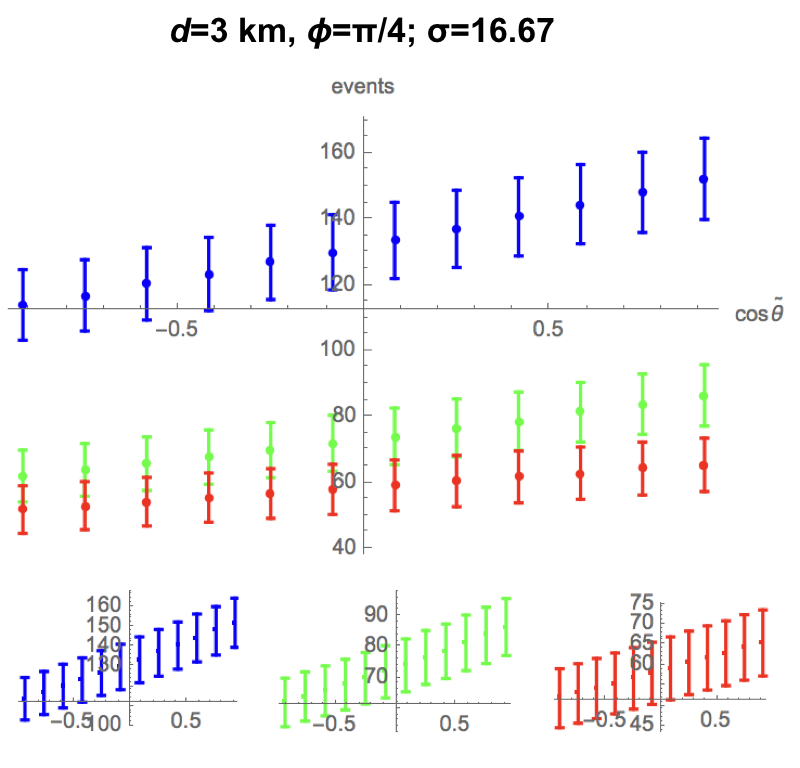}
		\end{subfigure}\\
		\begin{subfigure}{.27\textwidth}
		\includegraphics[width=\linewidth]{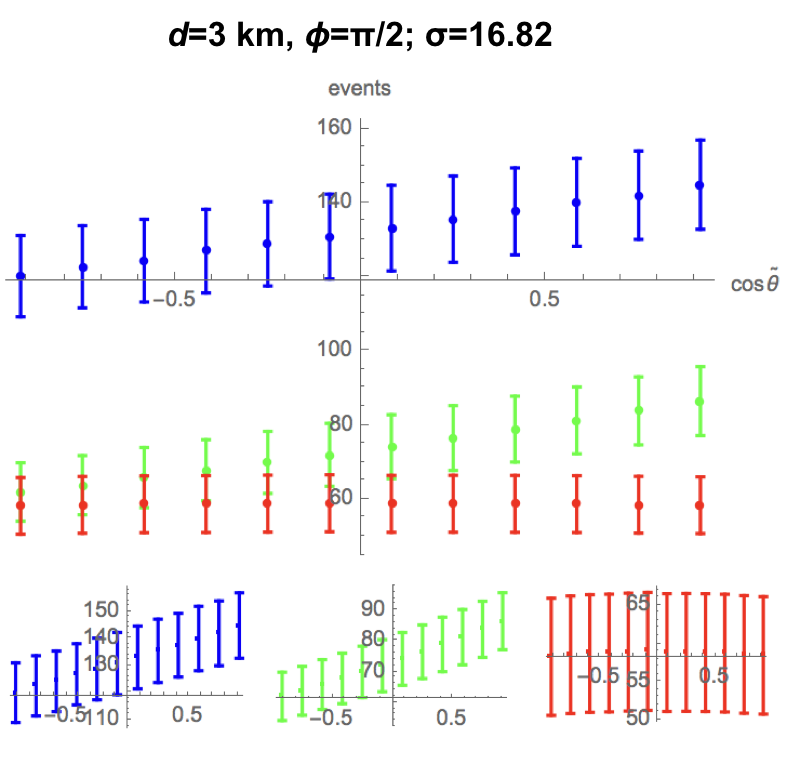}
		\end{subfigure}\\
		\begin{subfigure}{.27\textwidth}
		\includegraphics[width=\linewidth]{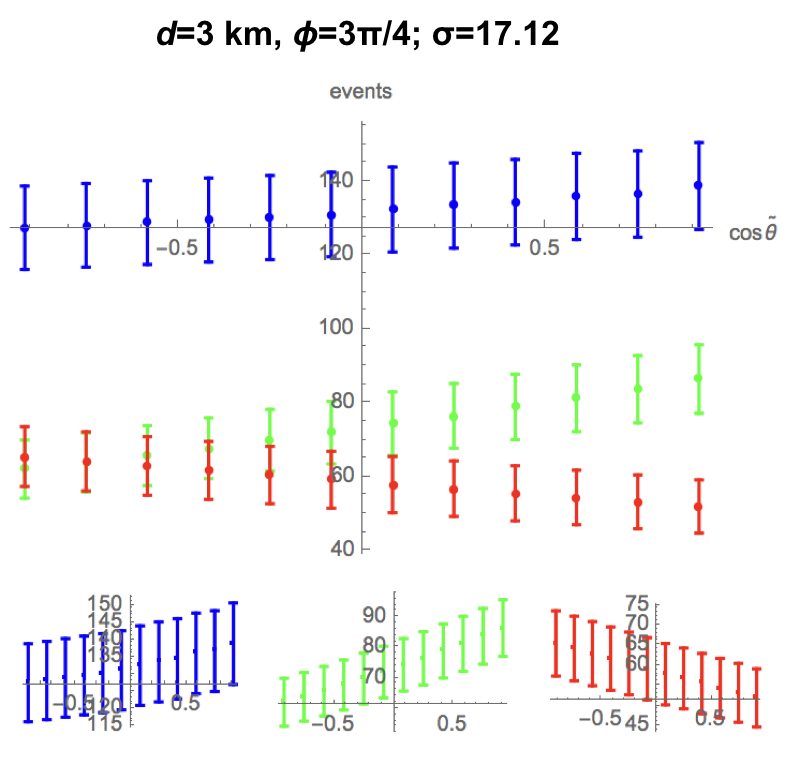}
		\end{subfigure}
\caption{Predicted reconstructed cosine histograms after one year monitoring two reactors for a range of azimuthal separations $\phi\in\{0,\pi/4,\pi/2,3\pi/4\}$, for an unknown-reactor standoff of $d=3\mathrm{\ km}$. Larger $\sigma$ values indicate a stronger significance to detect the unknown reactor (reject the null hypothesis). We plot by to the same method as described in Figure~6, which shows the comparable result for $\phi=\pi$.}
\end{figure}
\begin{figure}
		\begin{subfigure}{.29\textwidth}
		\includegraphics[width=\linewidth]{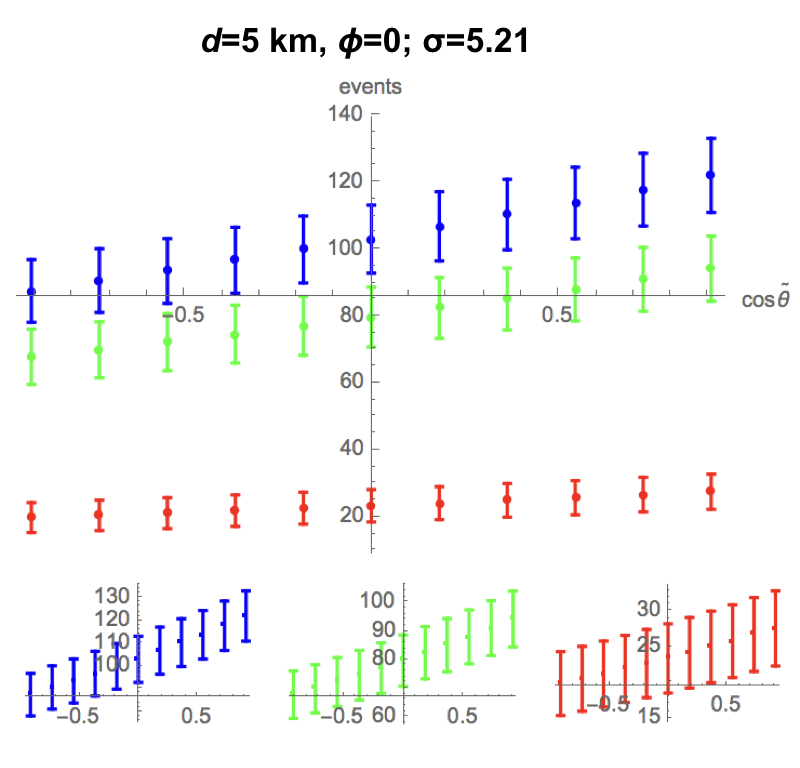}
		\end{subfigure}\\
		\begin{subfigure}{.29\textwidth}
		\includegraphics[width=\linewidth]{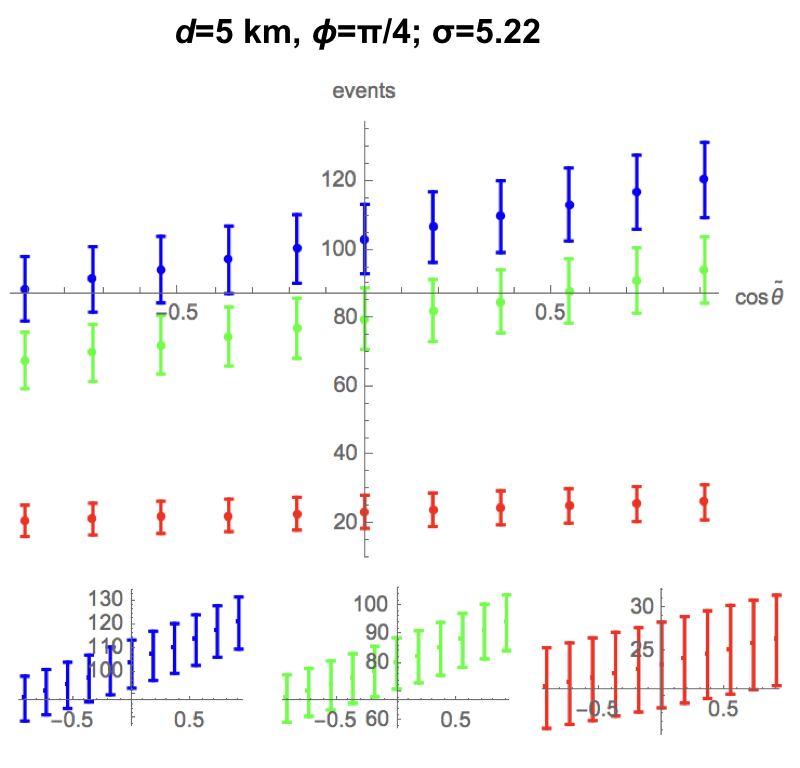}
		\end{subfigure}\\
		\begin{subfigure}{.29\textwidth}
		\includegraphics[width=\linewidth]{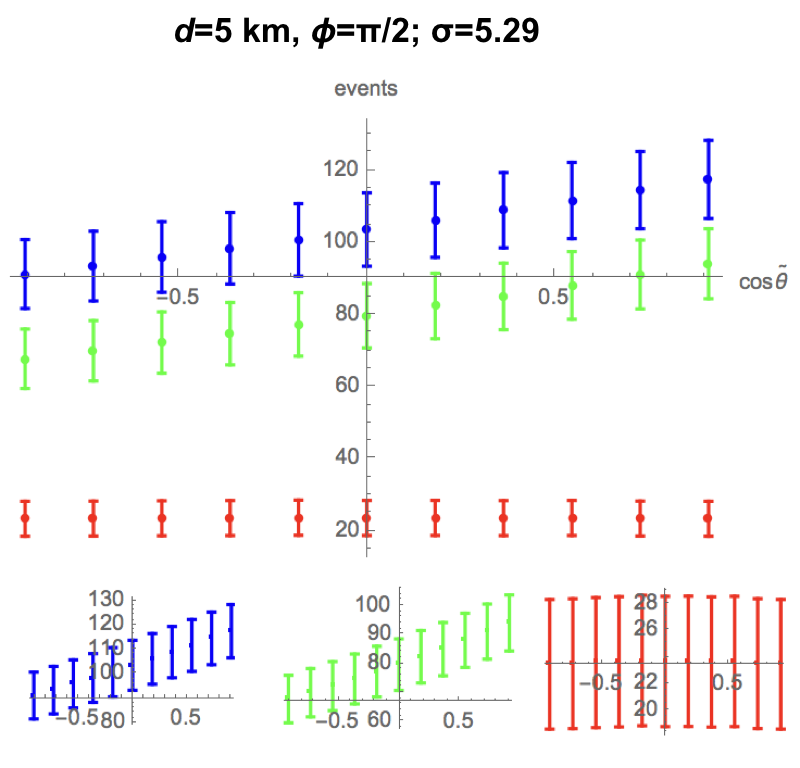}
		\end{subfigure}\\
		\begin{subfigure}{.29\textwidth}
		\includegraphics[width=\linewidth]{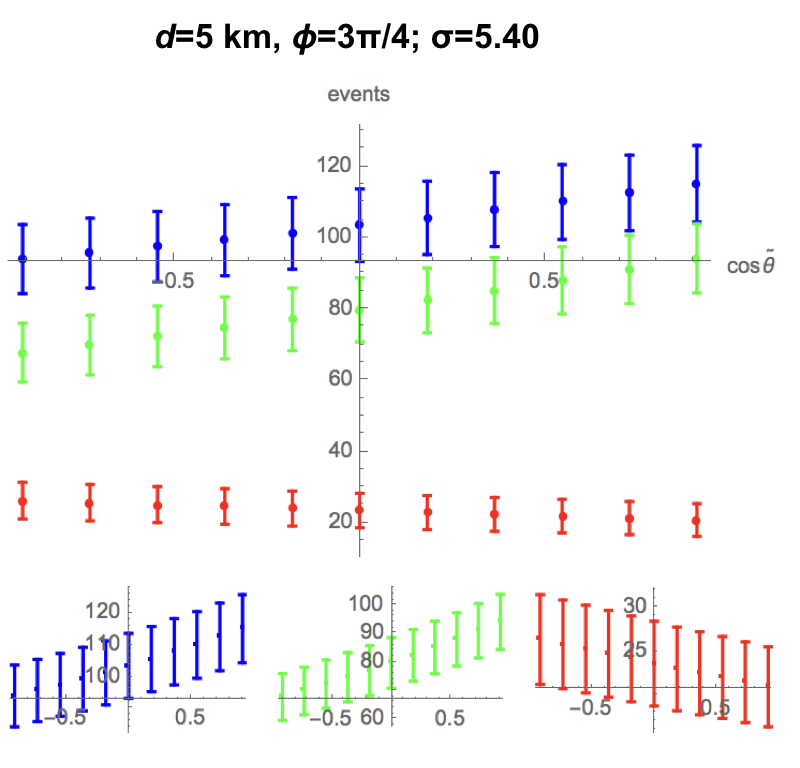}
		\end{subfigure}
\caption{Predicted reconstructed cosine histograms after one year monitoring two reactors for a range of azimuthal separations $\phi\in\{0,\pi/4,\pi/2,3\pi/4\}$, for an unknown-reactor standoff of $d=5\mathrm{\ km}$. Larger $\sigma$ values indicate a stronger significance to detect the unknown reactor (reject the null hypothesis). We plot by to the same method as described in Figure~6, which shows the comparable result for $\phi=\pi$.}
\end{figure}
\FloatBarrier
\bibliography{twoReactor}
\end{document}